# Broadside High Index Dielectric Beam Routing

[1]Arnab Dewanjee, [1]Stewart Aitchison [1]Mo. Mojahedi
[1]Department of Electrical and Computer Engineering,
University of Toronto, 40 King's College road, Ontario-M5S3G4, Canada
Author email: arnab.dewanjee@mail.utoronto.ca

**Abstract:** *We propose a technique to control the beam direction of a dielectric optical micro-prism structure by properly designing its geometry. We show that by selectively choosing the slope of a high index micro prism, the direction of the radiated beam from the tip can be controlled in the broadside direction of the prism structure. We propose an easy fabrication method to integrate such broadside beam routers on a Si platform. We also propose a demonstrate a technique to build the beam router on an elastic PDMS platform which demonstrates stress induced beam scanning in the broadside This specific feature of beam direction control can find application in free space optical interconnects, plasmonics, solar cells, detectors, optical sensing etc.*

1. **Introduction**

Delivering light to the nanophotonic components of an optical system by controlling the spatial dynamics of light has been a topic of wide interest in the field of nanophotonics. Plasmonic nano-antenna[1], multilayer metal-dielectric stacks [2] or metal slot waveguides [3] have been designed as techniques to confine light in subwavelength dimensions for diverse applications in nanophotonics. Stockman *et al.* have proposed a technique of localizing the optical energy in a small part of a nanosystem by means of coherent control [4]. Maksymov *et al.* demonstrated optical beam steering by 3D Yagi-Uda like metal nano-antenna array [5]. Volpe *et al.* has proposed a subwavelength spatial phase control method to reconfigure the optical near field distribution near plasmonic nano structures [6]. Furthermore, the control over the flow of optical energy have been proved to be useful in diverse fields like photo-detection [7,8] and photovoltaics [9], nano-imaging [10] and single photon emission [11]. In majority of these applications the antennae were constructed based on the plasmonic properties of metal. The collective electron oscillation influenced by the interaction of an electromagnetic wave with metal introduces the antenna-like properties to the metallic nanostructures (i.e. nanoparticles or bow tie structures). The control over the emission direction thus can be achieved by the controlling optical phase or the geometric parameters of the metal nanostructures. But one drawback of optical field control using metallic nanostructures is the Joule heating of metal and the concomitant optical loss. Dielectric beam routing structures have been proposed by several researchers in literature. Xue *et al.* have studied intra-chip free space communication using dielectric overhead micro lenses and mirrors to establish communication between different parts of the chip on a GaAs chip integrated with silicon photonics [12]. Gu *et al.* have demonstrated use of high index polymer surface couplers and integrated GaAs/AlAs multiple quantum well devices directly onto Si chip [13]. Chronis *et al.* have utilized the angled Si wet etch property of KOH to deflect a vertical beam to excite SPP by total internal reflection [14]. Also, Chao *et al.* has used total internal reflection at the facet of an angle polished fiber to couple to Si photonic chip through a grating coupler [15].

Here, in this paper, we show that by controlling the slope of a high index dielectric non adiabatic tapered micro prism structure, a control over the direction of the beam released from the prism can be directed to the broadside direction. Furthermore, a control over the beam direction can be achieved by controlling the slope of the prism structure. We also present an experimental demonstration of the broadside beam routing by a Si micro-prism on a silicon wafer. Furthermore, we have used Polydimethylsiloxane (PDMS) to build a stretchable micro-prism which facilitates a beam scanning in the broadside direction at the 1550 nm wavelength regime upon application of lateral stress. Since, PDMS is also a transparent high index material (as compared to air) in the visible wavelengths, such beam scanning can also be achieved in the visible regime. This technique of beam direction control can be very useful in different applications of nanophotonics such as free space interconnects, surface plasmon polariton (SPP) mode excitation in the visible wavelengths, guiding of light in the active layers of photovoltaic cells, plasmonic sensors, nonlinear signal conversion and photon detection.

The rest of the paper is organized as follows: in the next section we discuss in detail about the beam routing mechanism utilizing the total internal reflection inside a high index micro-prism. The next two sections (section 3 and section 4) discuss about the fabrication and measurement of the Si micro-prism for broadside beam routing. Then in section 5 and 6 we discuss about the fabrication process of the PDMS beam scanning micro-prism and the measurements of the stress induced beam scanning using the PDMS micro-prism.

## 2. Beam Routing Mechanism

In a prism submerged in a low index medium (i.e. air), as light is incident at one of its lateral facets from inside its high refractive index medium at an angle larger than the critical angle (between the prism material and air), a total internal reflection occurs and light is reflected back into the high index medium. According to Snell's law, the angle of reflection equals to the angle of incidence. The final direction of the reflected beam upon incidence on any of the other facets of the prism is dependent on the geometry and the refractive index of the prism as well as that of the surrounding medium. Since we aim to device a beam routing system compatible to integrated Si photonics, we choose a Si micro-prism structure with a sidewall angle of $54.7^0$ as depicted in Figure 1. The reason behind the choice of the inclination angle is due to the fact that a KOH wet etch of a masked <100> Si wafer creates a sidewall angle of $54.7^0$ as the Si is etched down from the edge of a mask pattern. Adopting this fabrication process to achieve a fixed prism angle of $54.7^0$ as shown in Figure 1 helps get rid of any grey scale lithography to manually control the inclination of the prism sidewall. Figure 1 depicts the routing of a beam upon total internal reflection from one of the slanted facets of a prism with the prism angle of $54.7^0$. Due to the $54.7^0$ inclination, the angle of incidence of the beam on the first slanted wall of the prism (shown by the yellow arrow in Figure 1) is $54.7^0$ and the angle of incidence of the second slanted wall of the prism (depicted by the angle ABO in Figure 1) is $15.9^0$. The angle that the beam is released from the prism ($\theta_{rl-air}$ in Figure 1) depends on the refractive index of the surrounding medium as well as the prism angle. Considering the surrounding medium as air, the angle that the beam is released from the prism ($\theta_{out-air}$) with respect to the normal direction of the surface can be calculated from the Snell's law to be $70.46^0$. This released beam makes an angle ($\theta_{rl-air}$) of $35.16^0$ with the horizontal direction as shown in Figure 1. Changing the surrounding medium to be oxide (refractive index 1.44) the released beam can be made almost horizontal ($\theta_{rl-oxide} = 6.2^0$).

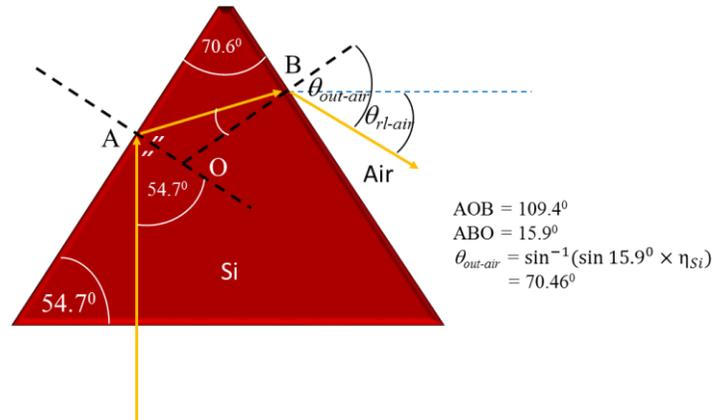

**Figure 1. Ray Diagram of beam routing in a Si prism of $54.7^0$ slope**

In the practical case of such a Si micro-prism being illuminated from the bottom (i.e. in an integrated photonic chip), the beam diameter (the mode field diameter) is typically comparable to the base of the Si prism. Therefore, although the routing mechanism still remains the similar as discussed above, directionality of the released beam depends on the position of the input beam with respect to the center of the base of the micro-prism. We used a commercial-grade simulator based on the finite-difference time-domain (FDTD) method to simulate the beam routing through a micro Si prism as discussed above [16]. Figure 2(a) & Figure 2(b) show the power density plot of unidirectional beam routing using a Si micro-prism submerged in air and $SiO_2$ respectively. As evident from the figures, a shifted beam towards the left results in the majority of the power in the beam to be routed towards the opposite (right) side of the prism and is released into the surrounding medium in the direction determined by the Snell's law. Figure 2(c) shows the power density distribution of a beam traversing through such a Si micro-prism with a centered beam configuration with $SiO_2$ as the surrounding medium. In this case, the finite width of the beam results in being incident partially on both the slanted walls of the micro-prisms. As a result, the beam is partially directed towards both the sides of the prism following the same rules of reflection and refractions and gets released in the surrounding lower indexed medium.

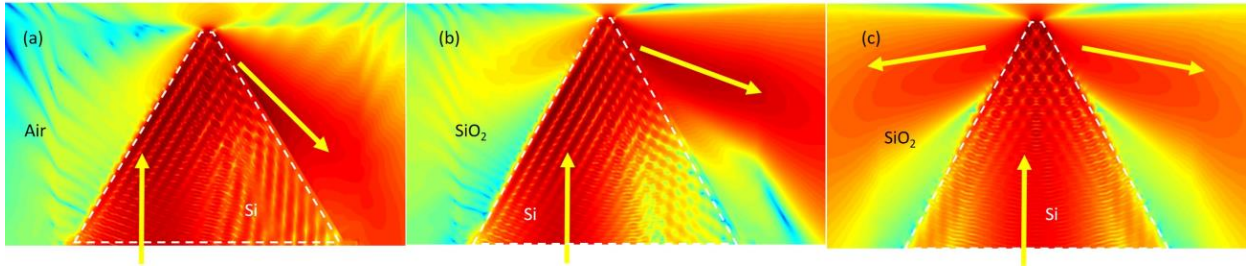

**Figure 2.** Logarithmic power density profile of beam routing from the Si micro-prism for three cases: (a) submerged in air and illumination shifted towards left for directional release of beam towards right, (b) submerged in SiO2 and illumination shifted towards right, and (c) submerged in SiO$_2$ and centered illumination for bidirectional release of beam towards right. The illumination direction from the bottom and the direction of the beam released from the prism is shown by the yellow arrows.

## 3. Fabrication

The fabrication of the Si micro-prism structure is fairly straight forward and require a single step lithographic process. First, a <100> Si wafer is thermally oxidized (300 nm) in a Bruce Technologies Inc. furnace for 18 min. The ma-N 2400 negative resist was spun on the oxide layer and the resist was patterned as a single 500 nm wide and 5 mm long line using the Vistec EBPG5000+ electron beam (Ebeam) lithography unit at the Toronto Nano-Fabrication Center TNFC) (step 3 in Figure 3). This is the only lithographic step in making the Si micro-prisms. After development of the exposed resist, an oxide etch was carried out using the Oxford Instruments PlasmaPro Estrelas100 DRIE System. As the next step we prepared a 45% KOH solution and heated it up to a stabilized temperature of $80^0$ C. Then the <100> Si sample with patterned oxide mask was submerged into the 45% KOH solution on a Teflon holder for 25 mins. The etch rate of the prepared solution was slightly less than 1 μm/min; therefore, after the 25 min etch we obtained a Si prism of 20 μm height (translates to a base size of 28.32 μm) as depicted in the schematic fabrication process diagram in Figure 3. For reducing diffused reflection at the back surface of the sample while coupling light during the characterization experiment (described in section 4), we carried out chemical-mechanical polishing of the back plane to achieve a smooth surface (~20 nm average roughness).

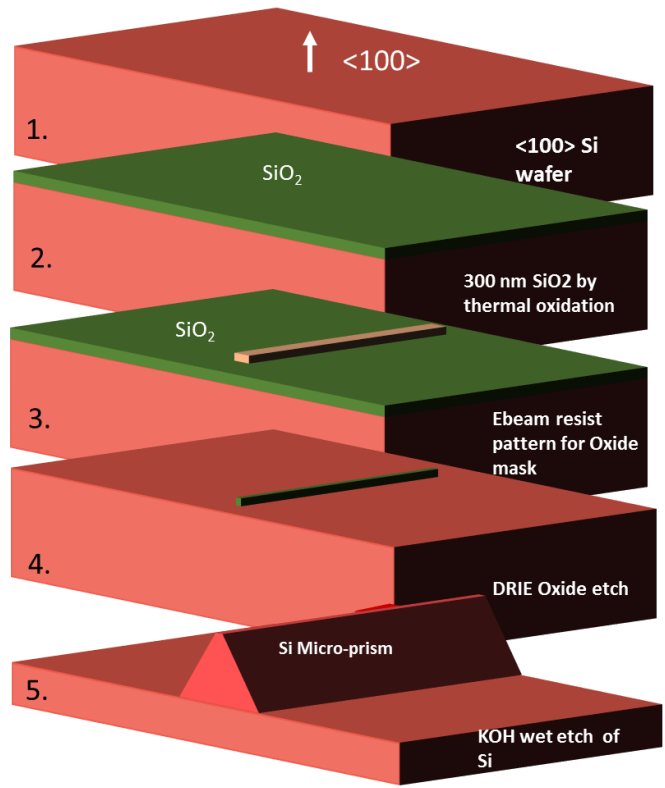

**Figure 3.** Fabrication process of the Si micro prism. The backside polishing step using CPM at the end is not shown in the figure.

### 4. Beam routing experiment

For the characterization of the fabricated sample, we built a radiation pattern measurement setup with the sample attached at the center of a rotational stage (Figure 4(b)). Since the wavelength of interest for this experiment is the telecom wavelength (1550 nm), we used a JDS Uniphase SW515101 Tunable Laser diode source tuned to emit at a center wavelength of 1550 nm. From the source, light was carried through a single mode fiber to the characterization table and converted to a free space Gaussian beam using a fiber collimator lens. The Gaussian beam was focused at the base of the Si prism as shown in Figure 4(a) using an objective lens (Newport F-L40B). The beam was focused following the shifted beam topology as mentioned in section 2 to achieve unidirectional beam routing. Then the radiation pattern was measured using the Newport 818-IR detector attached to the rotating arm of the rotation stage (Figure 4(b)).

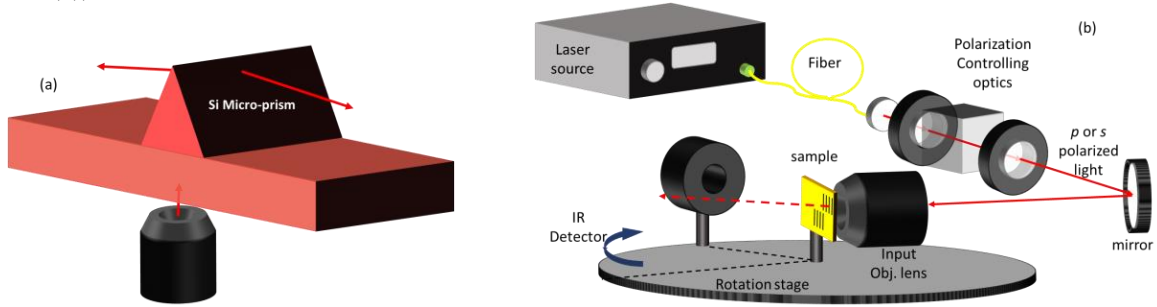

**Figure 4.** (a) Schematic diagram of the sample with the Si micro-prism. The red arrows indicate the routing of the beam. (b) Experimental setup for radiation pattern characterization.

The red dashed curve in Figure 5 shows the measured radiation pattern of the Si micro-prism and compares to the simulated results plotted by the green solid curve. It is evident from the figure that the measured radiation pattern closely matches the theoretically predicted pattern.

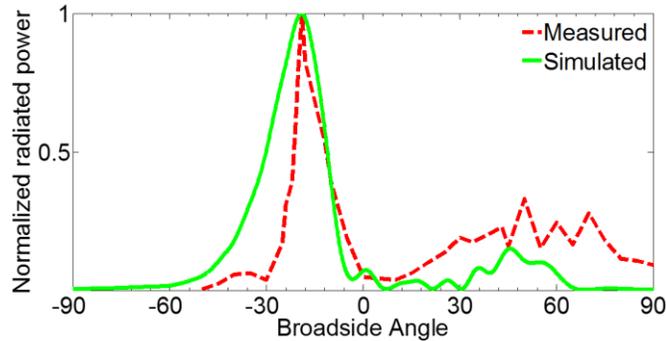

**Figure 5.** Comparison of simulated (green solid) and measured (red dashed) radiatoin pattern in the broadside direction using the Si micro-prism.

### 5. PDMS Micro-Prism for Beam Steering

As described in section 3, for the broadside beam routing through the micro-prism, the angle of release ($\theta_{rl\text{-}air}$ in Figure 1) of the out-coming beam is solely dependent on the slope of the sidewall of the deflecting prism. Therefore, a control over the direction of the released beam is achievable if the slope of the micro prism can be arbitrarily controlled. For such purpose, naturally, a solid Si micro-prism like the one described in the previous sections is not an ideal choice. Instead, an elastic material with a relatively higher index than air, for which, the absorption of light is negligible would be appropriate for the purpose. Cured PDMS is an elastic polymer and has a refractive index of 1.4 and is transparent at visible and near infrared wavelengths. Hence, we chose PDMS to be the material to fabricate our beam scanning micro-prism structure effective in both infrared and visible wavelengths. Our aim was to apply lateral stress on a PDMS micro-prism and change the sidewall slope which can facilitate scanning of the deflected beam in the broadside direction.

To fabricate the PDMS mico-prism we first used a Si substrate to etch a V shaped groove of 20 μm depth which is then used as a mold to shape the surface of a PDMS film using a pattern transfer method. Figure 6 shows the schematic of the steps of fabricating the PDMS micro-prism structure on a PDMS film. Once again we started the fabrication process by acquiring a piece of a thermally oxidized <100> Si wafer (as mentioned in section 3). The size of the oxidized Si sample was of a regular microscope glass slide. We then spun a positive resist (ZEP 520) compatible to electron beam lithography on the oxidized surface of the sample. Next, using the same Vistec EBPG5000+ Ebeam unit at the TNFC facilities we patterned a rectangular opening in the resist layer with the dimensions of 28.32 μm width and 5 mm length. As the next step, we did a DRIE etching (PlasmaPro Estrelas100 DRIE System) such that the oxide mask layer on the Si sample gets etched out in the region of the patterned opening in the resist layer. Then we ran the wet KOH etching of the sample with a 45% KOH solution at $80^0$ C temperature for ~ 25 mins. The KOH etch starts etching the exposed Si in the opening area of the oxide mask on top of the <100> Si substrate creating the side walls at the characteristic $54.7^0$ angle. At this point (step 5 of Figure 6) we got our mold ready for patterning the PDMS.

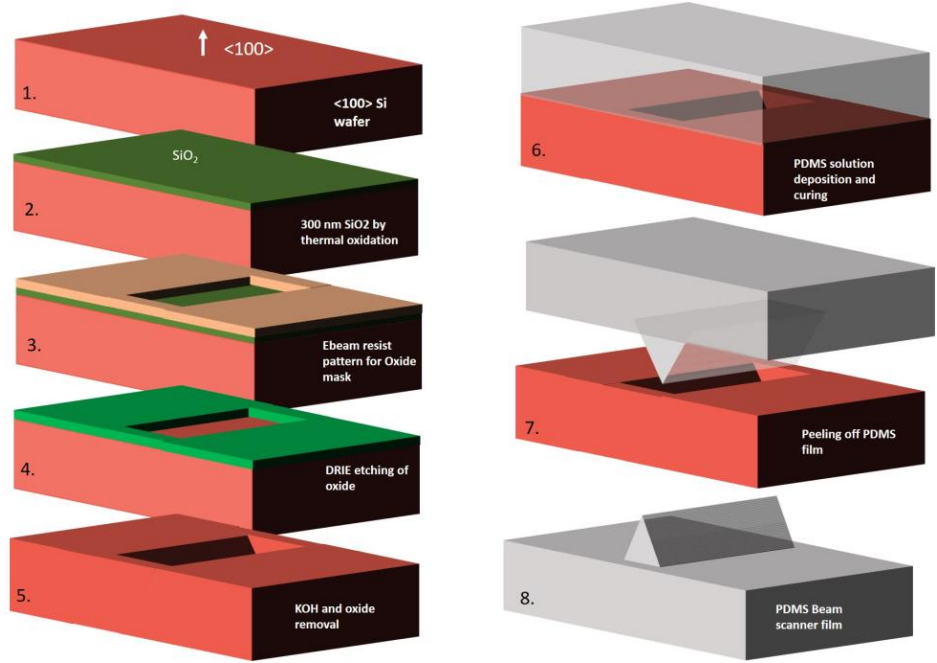

**Figure 6. Fabrication steps of the PDMS beam scanning film.**

We used a SYLGARD 184 SILICONE ELASTOMER kit from Dow Corning Corporation to prepare a flexible PDMS film with the micro-prism on the top surface of the film. We prepared a 10 to 1 mixture of the two liquids in the package as suggested by the Dow Corning data sheet and applied the viscous mixture on top of the prepared Si substrate containing the V groove in a petri dish. We prepared enough mixture so that the thickness of the liquid mixture after applying on the Si substrate was ~ 1mm. Then we put the petri dish along with the applied PDMS and the Si mold in a vacuum chamber for 45 minutes to get rid of the bubbles from inside the liquid PDMS. After the vacuum treatment, we put the petri dish in an oven at a temperature of $70^0$ for 1.5 hours for the curing of the PDMS. After curing the PDMS solidified and created a transparent elastic film on top of the Si substrate which contained, on its surface that is in contact with the top surface of the Si mold, the inverse of the features initially patterned on the top surface of the Si substrate. Next, with a sharp scalpel we cut the thick PDMS film around the Si substrate. Then we separated the substrate mold along with the PDMS film on top of it from the petri dish. At this point we cut the PDMS film again in a rectangular shape along the edges of the Si mold such that the cut-lines fall inside the edge of the Si mold. Then, starting from a corner, we carefully peeled off the cured PDMS film from top of the Si substrate. Thus we got the PDMS film that contains a micro-prism on its top surface embossed by the v-shaped groove on the top surface of the Si substrate mold (steps 7 and 8 in Figure 6).

### 6. Beam scanning measurement

Figure 7(b) shows the experimental setup for the measurement of the stress induced beam scanning using the PDMS micro-prism sample. We used a PELCO Large Universal Vise Clamp as a stretchable center stage to hold the stretchable PDMS film perpendicular to the input beam in the measurement setup depicted in Figure 7(b). The PELCO stage is capable of applying lateral stress on the PDMS film. Such lateral stress expands the base size of the PDMS micro-prism structure as depicted in Figure 7(a). Such stress induced expansion of the base of the PDMS micro-prism changes the inclination of the sidewalls of the prisms. From the discussion in section 2, Such change in the sidewall inclination results in a change in the direction of the deflected beam ($\theta_{rl\text{-}air}$) by the micro-prism. Due to the physical structure of the PELCO clamp used to introduce stress on the PDMS sample, it was required to use a single mode fiber to deliver the light from the JDS Uniphase diode laser source to the base of the PDMS microprism coupling light through the objective lens as shown in Figure 4-b. The PELCO clamp stage allowed to gradually increase the stress on the PDMS film from zero deformation to a maximum limit before tearing off the PDMS film.

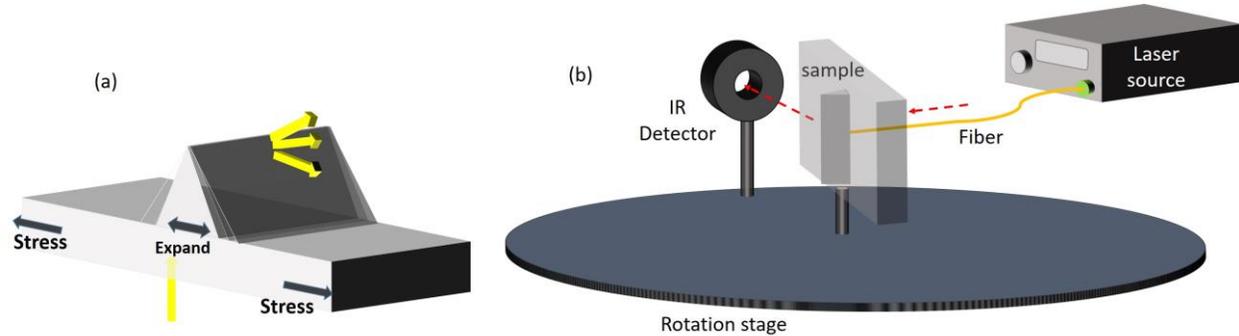

**Figure 7. a) Schematic of the PDMS film with the microprism being expanded with different lateral stresses. The yellow arrows depict the input beam and the change of deflected beam directions due to stress. b) schematic of the beam scanning measurement setup.**

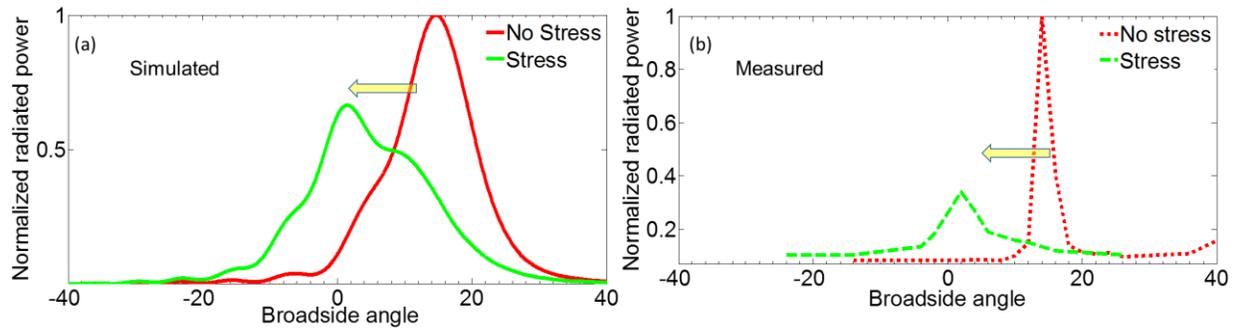

**Figure 8. a) Simulated radiation pattern of PDMS micro-prism. Red solid curve shows the radiation pattern for a PDMS micro-prism with prism angle of 54.70 while green solid curve shows the changed radiation pattern due to stress. b) Measured radiation pattern of the fabricated PDMS micro-prism corresponding to the stress and no stress case in (a). the prism angle for no stress condition is $54.7^0$ while with stress the angle reduced to $51.6^0$ due to enlargement of the prism base as a result of stress.**

Figure 8 presents the change in the deflected beam direction as a function of the broadside angle due to the expansion of the prism base as a result of the change in applied lateral stress. The solid curves represent the radiation patterns calculated by the simulation software [16] while the dashed curves represent the corresponding measured data. The evident conformity between the simulated and the measured radiation pattern confirms effectivity of the mechanical beam scanning concept using the mechanically stretchable PDMS film.

## 7. Conclusion

In this paper, we introduced a concept of beam routing using an integrated micro-prism on a <100> Si chip utilizing the total internal reflection inside a Si prism. We fabricated the proposed Si micro prism and demonstrated beam routing in the broadside direction. We then extended this beam routing concept to a flexible PDMS platform and demonstrated beam scanning in the broadside direction by applying lateral stress on the film containing a PDMS micro-prism on its surface. The proposed beam routing scheme can find potential applications in integrated photonics and act as a vertical via for multilayer photonic circuitry. The demonstrated PDMS beam scanning technique can be useful in exciting surface plasmon polaritons at a metal/dielectric interface in visible wavelengths with significantly high efficiencies.